\newif\ifjournal\journalfalse

\documentclass[preprint,showpacs,preprintnumbers,amsmath,amssymb]{revtex4}

\usepackage{graphicx}

\newcommand{\newrev}[1]{{#1}}

\newcommand{\grad}{\nabla}
\renewcommand{\div}{\nabla \cdot}
\newcommand{\rot}{\nabla \times}
\newcommand{\pp}[2]{\frac{\partial #1}{\partial #2}}
\newcommand{\Alfven}{Alfv\'{e}n }

\def\grl{{\itshape Geophys. Res. Lett.} }
\def\jgr{{\itshape J. Geophys. Res.} }
\def\apj{{\itshape Astrophys. J.} }
\def\prl{{\itshape Phys. Rev. Lett.} }
\def\pop{{\itshape Phys. Plasmas} }
\def\aap{{\itshape Astron. Astrophys.} }
\def\mnras{{\itshape Monthly Notices of the RAS} }
\def\nature{{\itshape Nature} }

\begin{document}

\title{The role of the Weibel instability at the reconnection jet front \\
in relativistic pair plasma reconnection}

\author{S. Zenitani}
\author{M. Hesse}
\affiliation{
NASA Goddard Space Flight Center, Greenbelt, MD 20771, USA
}
\email{zenitani@lssp-mail.gsfc.nasa.gov}

\ifjournal
\else
\date{submitted to \it{Physics of Plasmas}}
\fi

\begin{abstract}
The role of the Weibel instability is investigated for the first time
in the context of the large-scale magnetic reconnection problem.
A late-time evolution of
magnetic reconnection in relativistic pair plasmas
is demonstrated by particle-in-cell (PIC) simulations.
In the outflow regions,
powerful reconnection jet piles up the magnetic fields
and then a tangential discontinuity appears there.
Further downstream, it is found that
the two-dimensional extension of
the relativistic Weibel instability
generates electro-magnetic fields,
which are comparable to the anti-parallel or piled-up fields.
In a microscopic viewpoint, 
the instability allows plasma's
multiple interactions with the discontinuity.
In a macroscopic viewpoint, 
the instability leads to rapid expansion of the current sheet
and then the reconnection jet front further propagates into the downstream.
Possible application to the three-dimensional case is briefly discussed.
\end{abstract}

\maketitle

\section{Introduction}

Magnetic reconnection is widely recognized as
a fundamental physical mechanism in collisionless plasmas.
Consuming the magnetic field energy in the inflow region, 
it releases the energy to the kinetic energy of plasma particles.
It is an effective engine for
magnetic dissipation, plasma heating, or particle acceleration.
Recently, the relativistic extension of magnetic reconnection
has received attention for its role
in various high-energy astrophysical places
--- active galactic nuclei \citep{lb97,dimatteo,birk01},
pulsars \citep{coro90,lyu01,kirk03}, gamma ray bursts \citep{dr02,drs02},
and magnetars \citep{thom95,lyut03a}. 
The mechanism of relativistic reconnection still remains unclear
as well as the conventional non-relativistic reconnection,
but recent kinetic simulations start to reveal its features.
It is demonstrated that
relativistic pair plasma reconnection is a powerful particle accelerator
\citep{zeni01,zeni05b,zeni07a,zeni07b,claus04},
and that it keeps fast reconnection rate \citep{bessho07,hesse07}
despite the lack of the Hall physics \citep{birn01}.
In principle, magnetic reconnection is a relatively large scale process ---
the typical speed of reconnection jet is
the \Alfven velocity $V_A$ or the light speed $\sim c$,
and the typical time scale is several tens of characteristic time scale;
$\lambda/V_A$ or $\lambda/c$,
where $\lambda$ is the typical spatial scale of the field reversal.

On the other hand,
in the context of gamma ray bursts \citep{piran,med99}
or extra galactic jets,
the Weibel-type two-stream instability
in relativistic plasmas has attracted recent attention, too.
The Weibel instability \citep{weibel} is an electromagnetic instability,
that arises from plasma anisotropy.
Since it quickly generates magnetic fields,
it is a likely origin of magnetic field
in the synchrotron source
near collisionless shocks or near relativistic jet fronts.
Series of PIC simulations
\citep{kazi98,silva03,nishikawa03,fre04,hed04,claus05,nishikawa06}
successfully demonstrate the magnetic generation
via jet penetration or plasma shell collision
in a weakly or nonmagnetized plasma.
The Weibel magnetic structure evolves into long-durated
filament-like magnetic structure,
whose energy is approximately 10\% of equi-partition energy.
In-situ particle acceleration is also reported,
but its detailed mechanism still remains unclear \citep{kazi98,hed04,nishikawa06}.
On the theoretical side, the conventional Weibel instability,
which propagates into the transverse direction from plasma anisotropy,
has been extended to relativistic temperatures \citep{yoon87,yang93,yoon07}.
Meanwhile, its two-dimensional extension,
the electromagnetic counter streaming instability, has been studied
in relativistic counter streaming conditions \citep{cal97,kazi98,saito}.
In general, the Weibel instabilities are microscale phenomena,
whose scales are characterized by the plasma frequency $\omega_p=[4\pi ne^2/m]^{1/2}$; $\omega_p^{-1}$ in time and by $(ck/\omega_p)$ in space.

In the context of magnetic reconnection,
since magnetic reconnection expels powerful outflow jets 
from the reconnecting region,
it is quite possible that the jets interact with pre-existing plasmas,
and then excite an anisotropy-driven instability. 
In fact, \citet{dau07} reported generation of out-of-plane magnetic field
via firehose type instability in their non-relativistic pair plasma reconnection,
although its role in reconnection remains unclear.

In the present paper, we study
the role of the Weibel instability in the reconnection context.
We carry out two-dimensional PIC simulations of
relativistic pair plasma reconnection, and 
we find that the relativistic counter-streaming Weibel instability
generates out-of-plane magnetic fields
in the downstream region of reconnection outflow.
We discuss the properties of the instability,
and then we investigate how the Weibel instability affects
micro- and macro physics of magnetic reconnection.
The paper consists of the following sections.
In section II we describe our simulation setup.
In section III we present the two-dimensional simulation results,
and then in section IV we investigate the properties of the instability.
In section V we discuss how the Weibel instability effects plasma dynamics,
both in microscopic particle motion and macroscopic reconnection structure.
The last section contains discussion and the summary.



\section{Simulation setup}

We carry out two-dimensional particle-in-cell (PIC) simulations
in a current sheet configuration.
As an initial condition,
we employ a relativistic extension of the Harris model
in GSM-like geometry.
The magnetic field, plasma density and
plasma distribution functions are described by
$\vec{B} = B_0 \tanh(z/\lambda) \hat{\bm{x}}$,
$d(z) = d_0 \cosh^{-2} (z/\lambda) = (\gamma_{\beta} n_0) \cosh^{-2} (z/\lambda)$ and
$f_{s} \propto d(z) \exp[-\gamma_\beta\{\varepsilon - \beta_s u_y\} / T ] $.
In the above equations,
$B_0$ is the magnitude of antiparallel magnetic field,
$\lambda$ is the typical thickness of the current sheet,
\newrev{$d_0$ is the lab-frame number density in the current sheet,
$n_0$ is the proper number density,}
the subscript $s$ denotes the species (`p' for positrons, `e' for electrons),
$\beta_{p} = -\beta_{e} = \beta$ is the dimensionless drift velocity,
$\gamma_\beta$ is the Lorentz factor for $\beta$ ($\gamma_\beta = [1-\beta^2]^{-1/2}$),
$\varepsilon$ is the particle energy,
$\vec{u}$ is the relativistic four velocity of $\vec{u}= [1-(\vec{v}/c)^2]^{-1/2} \cdot \vec{v}$
and $T$ is the proper temperature including the Boltzmann constant.
We set $T = mc^2$ and $\beta = 0.3$, respectively.
In addition, a uniform background plasma is added to the system
\newrev{in order to supply plasmas in the reconnection inflow region.}
Its number density and temperature are
$d_{bg}/d_0 =5\%$ and $T_{bg}/mc^2=0.1$, respectively.
\newrev{
In general, the velocity of the reconnection outflow jet is known to be
approximately the \Alfven velocity in the inflow region.
In the present study, we choose a low-density (5\%) background population
to obtain fast outflow.
Notice that the Harris model with uniform background plasmas
exactly satisfies an equilibrium.}

The system consists of $1568 (x) \times 768 (z)$ grids
and the typical scale of the current sheet $\lambda$ is set to 10 grids.
Since we consider periodic boundaries in the $x$ direction,
and since there are two current layers in the periodic $z$ direction,
the boundaries of the main simulation domain are located at
$x = \pm 76.8 \lambda$ and $z = \pm 19.2 \lambda$.
We use $7.5 \times 10^7$ super particles in this simulation.
One cell contains $6.6 \times 10^2$ particles
at the center of the current sheet. 
\newrev{
During the very early stage
we impose a small artificial electric field $\tilde{E_y}$ around
$(x, z) = (0,\pm 3\lambda)$.
The resultant $\vec{E}\times\vec{B}$ flow
compresses the current sheet,
and then it triggers reconnection
around the center of the main simulation domain.
The typical spatial ranges of the trigger field $\tilde{E_y}$ are
set to $(\Delta x,\Delta z) \sim (\pm 2\lambda,\pm \lambda)$.}
The trigger field soon vanishes after $t/\tau_c=(10-15)$,
where $\tau_c=\lambda/c$ is the light transit time.
We discuss the physics of reconnection
in the late stage of $t/\tau_c=(60-120)$,
which is not influenced by this initial perturbation.
These conditions are similar to
the author's previous study \citep{zeni07a}; 
but we use a larger simulation box
to discuss late-time structure without boundary effects.
We call this reference run `run A'. 
The total energy is conserved within an error of $0.1\%$
throughout the simulation run, after the initial trigger force vanishes.

\section{Results}

\begin{figure}[thp]
\begin{center}
\ifjournal
\includegraphics[width={\columnwidth},clip]{f1.eps}
\else
\includegraphics[width={\columnwidth},clip]{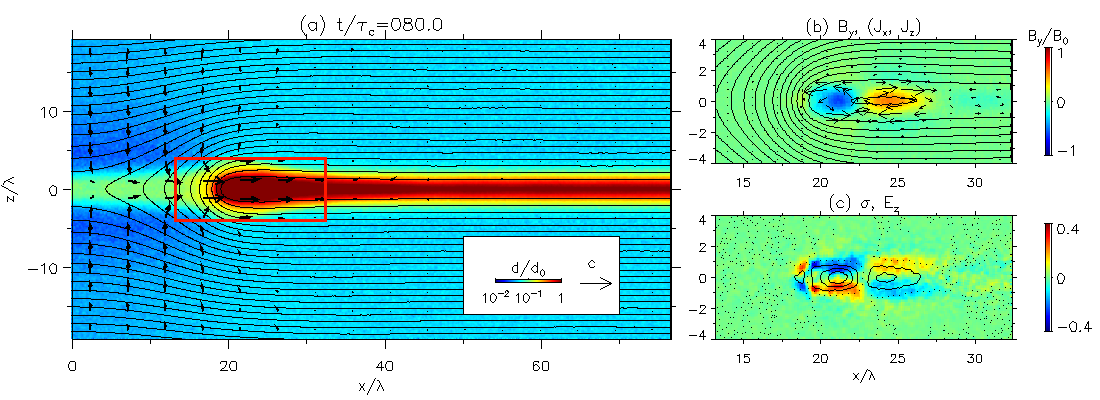}
\fi
\caption{(Color online)
\label{fig:snap}
(a) Snapshot of the right half of the main simulation domain.  Magnetic field lines (\textit{contour}), plasma density (\textit{color contour})
and plasma flow (\textit{arrows}).
(b) Out-of-plane field structure ($B_y$) in the selected region and the electric current system (\textit{arrows}) in the $xz$ plane.
(c) Charge distribution $\rho = [d_p-d_e]/d_0$ in color.  
The dotted line shows $E_z=0$ and the solid lines are contour of electric field $E_z$ with $\Delta E_z=0.1 B_0$.
}
\end{center}
\end{figure}

\begin{figure}[thp]
\begin{center}
\ifjournal
\includegraphics[width={\columnwidth},clip]{f2.eps}
\else
\includegraphics[width={\columnwidth},clip]{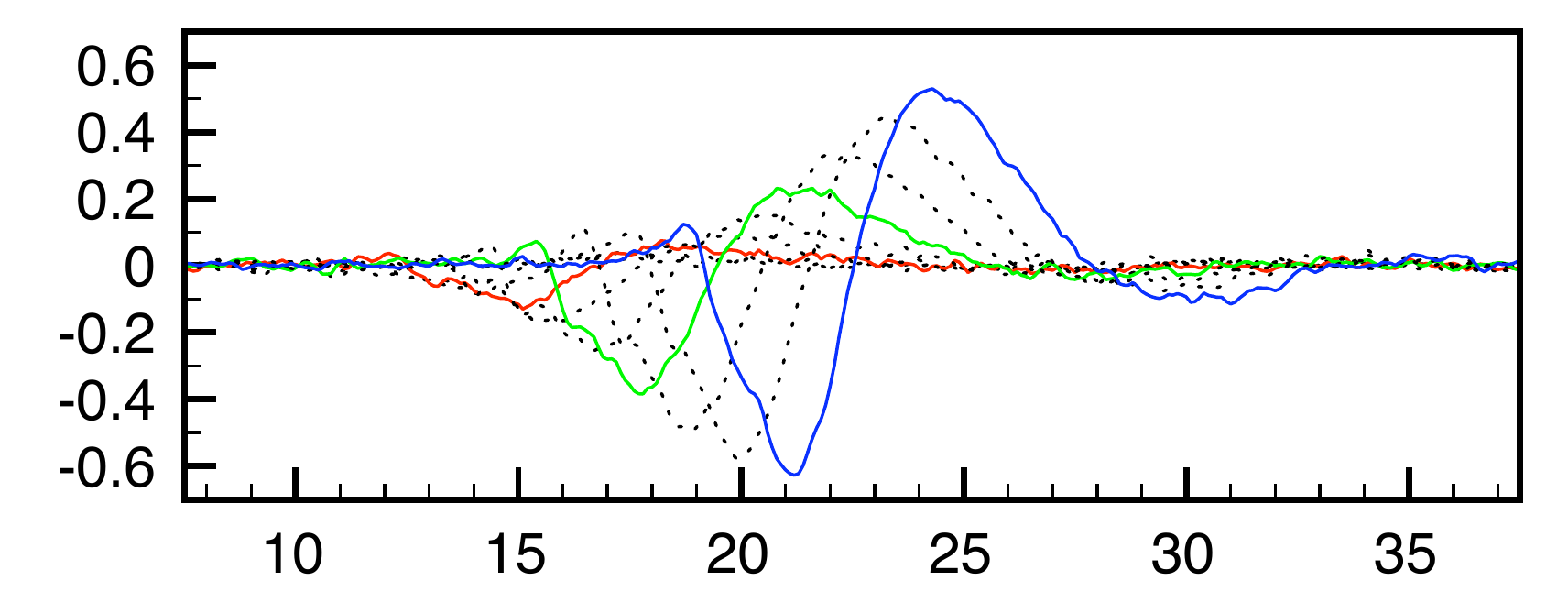}
\fi
\caption{(Color online)
\label{fig:linear}
Time development of out-of-plane magnetic field $B_y$
along the neutral plane ($z=0$).
Profiles at three stages ($t/\tau_c=68,74,80$) are indicated
by \newrev{solid} color lines.
}
\ifjournal
\includegraphics[width={0.8\columnwidth},clip]{f3.eps}
\else
\includegraphics[width={0.8\columnwidth},clip]{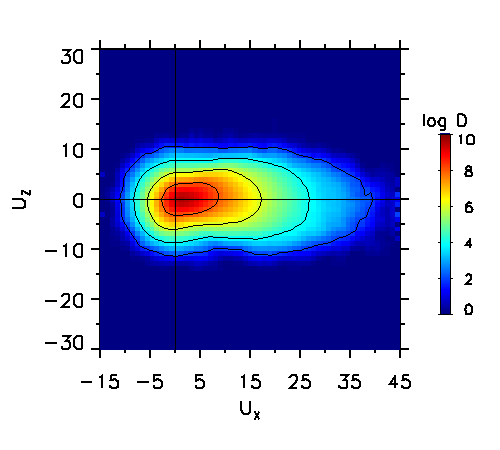}
\fi
\caption{(Color online)
\label{fig:aniso}
Plasma distribution function in the Weibel active region
($22 \le x/\lambda \le 26, -2 \le z/\lambda \le 2$) at $t/\tau_c=80$
in the $x$-$z$ four-velocity space, normalized by $c$.
}
\end{center}
\end{figure}

\newrev{Due to the initial perturbation,
magnetic reconnection takes place around the center of
the main simulation domain.
Magnetic field lines start to reconnect at $t/\tau_c\sim50$,
and outflow jets into the $\pm x$ directions appear}.
Figure \ref{fig:snap}\textit{a} show a snapshot of
the right half of the main simulation domain at $t/\tau_c=80$.
\newrev{The left half ($-76.8\le x/\lambda \le 0$) is not presented.
The reconnection outflow is well developed at this stage, and its}
speed is up to $\sim 0.7c$.
Along the neutral line,
magnetic fields are piled-up
in front of the dense plasma region of the current sheet
around $x/\lambda \sim 18$.
Its peak amplitude is $B_z/B_0\sim 1.5$, and
there is a relatively sharp boundary between
the pileup magnetic field and
the pre-existing dense plasma in the downstream.
This boundary is a tangential discontinuity (hereafter TD in short),
and we discuss ``upstream'' and ``downstream'' based on the TD
throughout this paper.
The propagation speed of the TD ($V_{TD}\sim 0.65c$) is
slightly slower than the average velocity of local plasmas.
The typical plasma density in the simulation frame is
$d/d_0\sim 0.1-0.2$ in the upstream,
$d/d_0\sim 2.5$ at the downstream side of the TD
and then it decreases to $d/d_0 \sim 1$ in the further downstream region.
Figure \ref{fig:snap}\textit{b} shows
the out-of-plane magnetic field ($B_y$) structure and
the relevant current system
in the reconnection outflow region,
which is indicated by the rectangle in Figure \ref{fig:snap}\textit{a}.
The characteristic structure of $B_y$ is observed, and
its maximum amplitude is $B_y\sim 0.6 B_0$.
In Figure \ref{fig:snap}\textit{c}
we observe charge separation at the same place and
the vertical $E_z$ structure
\newrev{(Hereafter the term ``vertical'' means the $z$ direction)}.
The $E_z$ explains both
the motional field for $B_y$
and the electrostatic field by the charge separation. 
The time development of the $B_y$ structure along the neutral plane
is presented in Figure \ref{fig:linear}.
These $B_y$ fields suddenly appear after $t/\tau_c\sim 64$ and then
they exponentially grow until they saturate after $t/\tau_c\sim 80$.
The instability looks like a convective mode, traveling into the $+x$-direction.
However, actually, it is nearly
non-convective purely-growing mode
in the frame of the plasma average flow. 
The linear growth rate measured by $B_y$ growth is
$\tau_c\omega_i \sim 1.7$-$1.8 \times 10^{-1}$ or 
$\omega_i/\Omega_p \sim 5.2$-$5.5 \times 10^{-2}$,
where $\Omega_p$ is the typical plasma frequency in the system.
The typical spatial scales are
$7\lambda$-$10\lambda$ ($x$) and $\sim 2\lambda$ ($z$).
Careful observation show that the instability has
a two-dimensional rectangular structure.
In Figure \ref{fig:snap}\textit{b},
we see the weak negative regions
on the upper side, on the lower side and on the right side
\newrev{[colored in light blue; $(x,z) \sim (24,\pm 1.5), (30,0)$ in unit of $\lambda$] }of
the characteristic positive region \newrev{[inner orange region; $\sim (24,0)$]}.
Similarly, weak positive regions are located in the vicinity of
the characteristic negative region \newrev{[inner blue region; $\sim (21,0)$]}. 
We find that these structures are generated by
the two-dimensional Weibel-type instability.
In this case, plasmas are highly anisotropic along the $x$-direction,
mainly because the TD pushes away the pre-existing plasmas,
and because the reconnection outflow jet penetrates into this region.
Therefore, the situation is similar to
jet injection \citep{cal97,nishikawa03} or 
relativistic counter-stream \citep{kazi98},
and magnetic generation near the shock \citep{med99} in pair plasmas.
The instability resides inside the current sheet,
where the plasma frequency is high.
In addition, the Weibel instability prefers an unmagnetized region,
and so an inner current sheet is an ideal place for the instability. 
The current structure and the charge separation structure 
indicates the nature of the Weibel-type activity.
As schematically explained in \citet{med99},
small $B_y$ fluctuation leads to
the $z$-displacement of $\pm x$-streaming plasmas,
and then the resultant \newrev{$x$-current structure $\pm \delta J_x$}
continues to enhance $\delta B_y$.
Thus, the Weibel instability generates magnetic field
which is perpendicular to the direction of the anisotropy,
and then it leads to the reduction of anisotropy.

Figure \ref{fig:aniso} shows the plasma distribution function
of $1.4 \times 10^5$ particles in the Weibel active region
($22 \le x/\lambda \le 26, -2 \le z/\lambda \le 2$) at $t/\tau_c=80$.
The left part of the distribution function is almost identical to
the initial distribution of pre-existing plasmas,
but the right part are highly elongated due to
both reflected plasmas and upstream-origin accelerated particles. 
\newrev{The average plasma velocity is $\langle v_x \rangle \sim 0.49c$,
where $\langle ~ \rangle$ means the average value among the particles.}
On the contrary, the plasma fluid velocity
(a Lorentz transformation velocity to the rest frame,
where the plasma momentum flow is zero)
is $\sim 0.65c \sim V_{TD}$.
The two velocities differ due to highly asymmetric plasma distribution.
An integrated plasma temperature is as follows;
\newrev{$(T'_x,T'_y,T'_z)=(\langle{m u'_x v'_x}\rangle, \langle{m u'_y v'_y}\rangle, \langle{ mu'_z v'_z}\rangle ) = mc^2 (2.3, 1.0, 1.5)$,
where $T'$, $m\vec{u}'$ and $\vec{v}'$ are
the temperature, momentum and velocity in the rest frame of plasmas.}
After the reconnection jets start from the $X$-type region,
the plasma anisotropy in the downstream region grows in time,
until the Weibel instability appears. 
The anisotropy stays at the same level after the instability occurs,
because $x$-momentum is continually supplied from the upstream side.

Snapshots of the field properties along the neutral line;
the pileup field $B_z$ (\textit{bold line}),
the reconnection electric field $E_y$ (\textit{dashed line}),
and the Weibel magnetic field $B_y$ (\textit{thin line}) 
are presented in Figures \ref{fig:stack}\textit{a-d}.
Note that the Weibel fields are observed
in the local frame of plasma average flow.
They propagate to the $+x$ direction and 
its speed is slightly slower than the speed of the TD ($V_{TD}$).
Therefore, sometimes the Weibel fields are caught up by the TD.
For example, the positive $B_y$ region
around $x/\lambda\sim 24$ at $t/\tau_c=80$
is nearly caught by the TD
around $x/\lambda\sim 34$ at $t/\tau_c=100$.
At the same time, new Weibel fields are continuously
generated in the further downstream region.
We discuss the late time development of the Weibel fields later.
The subpartition of Weibel field energy $(B_y^2+E_x^2+E_z^2)/8\pi$
to the local plasma kinetic energy saturates
around $8$-$12$\% in run A.

\section{Linear analysis}

In order to study the properties of the instability,
we have solved the dispersion relation
by linearizing relativistic fluid equations.
\newrev{In the Weibel region,
plasmas mainly consist of three different components;
(i) pre-existing plasmas in the Harris current sheet,
(ii) current sheet plasmas, which are reflected by the TD,
and (iii) upstream-origin plasmas, which are originally from
the reconnection inflow region.
Since we set a low plasma density
in the reconnection inflow region (5\% of the Harris current sheet),
the third population is relatively smaller than the other two.
Therefore, we employ the counter-streaming model of (i) and (ii)
in order to evaluate the Weibel instability.}
\newrev{We extend \citet{kazi98}'s fluid theory
for counter-streaming four fluids
(streaming/counter-streaming positrons and electrons),
which was originally developed by \citet{cal97}.
Although \citet{kazi98} ignored the plasma pressure effect,
it is here included.}
We employ the following relativistic fluid equations;
\begin{equation}\label{eq:fluid}
\frac{\gamma_{sa}^2}{c^2} (p_{sa} + e_{sa})
( \pp{}{t} + \vec{v}_{sa} \cdot \grad )
\vec{v_{sa}}
=
- \grad p_{sa} +
\gamma_{sa} q_{sa} n_{sa} 
\Big( \vec{E} +  \frac{\vec{v}_{sa}}{c} \times \vec{B}  \Big)
- \frac{\vec{v}_{sa}}{c^2} ( \gamma_{sa} q_{sa} n_{sa} \vec{E} \cdot \vec{v}_{sa} + \pp{p_{sa}}{t} ),
\end{equation}
where $p$ is isotropic plasma pressure,
$e$ is the fluid internal energy,
the subscript $a$ denotes two kind of streams
(`$1$' for streaming fluids, and `$2$' for counter-streaming fluids),
and $\gamma_{sa} = [ 1-(\vec{v}_{sa}/c)^{2} ]^{-1/2}$ is
the relevant Lorentz factor.
We also use the continuity equation and Maxwell equations
\begin{equation}\label{eq:cont}
\pp{}{t}(\gamma_{sa} n_{sa}) + \div(\gamma_{sa} n_{sa} \vec{v}_{sa} ) = 0,
\end{equation}
\begin{equation}\label{eq:rotb}
\rot \vec{B} = \frac{ 4\pi }{c} \sum_{s=e,p} \sum_{a=1,2} \gamma_{sa} q_{sa} n_{sa} \vec{v}_{sa} + \frac{1}{c} \pp{ \vec{E}}{t}
\end{equation}
\begin{equation}\label{eq:rote}
\rot \vec{E} =  - \frac{1}{c} \pp{ \vec{B}}{t}.
\end{equation}
We assume the adiabatic gas condition in order to close the equation
\begin{equation}\label{eq:p}
p_{sa} \propto n_{sa}^{\Gamma}
,~
e_{sa} = n_{sa} m c^2 + \frac{1}{\Gamma -1} p_{sa}
\end{equation}
where $\Gamma = 5/3$-$4/3$ is the polytropic index of adiabatic gas,
We consider a two dimensional perturbation
$\delta f \propto \delta f \exp(ik_x x + ik_z z - i \omega t)$,
where $\vec{k}=(k_x,k_z)$ is the wavevector
and $\omega$ is the complex frequency,
and then we linearize all equations for four fluids.
Then, we numerically calculate the growth rate (Im $\omega$)
for arbitrary wavevector $\vec{k}=(k_x,k_z)$ by solving a matrix problem.
For simplicity, the following assumptions are used;
\begin{eqnarray}
v_{p1}=v_{e1}=V_{TD},~v_{p2}=v_{e2}=-V_{TD}\\
n_{p1}=n_{p2}=n_{e1}=n_{e2}=n_0\\
p_{p1}=p_{p2}=p_{e1}=p_{e2}=n_0 mc^2 .
\end{eqnarray}
We assume that plasma density is homogeneous,
two counter-streams are symmetric,
the frame is set to the co-moving frame of the TD,
considering that the TD completely reflects the momentum of
pre-existing plasmas.  In the present case,
the simulation data shows
$\vec{k}=(k_x,k_z) \sim (\omega_p/c)(0.15,0.75)$
in the frame of interest.
Because of the complexity in the simulation system,
this analysis does not exactly describe the instability.
The density gradient of plasmas, the current sheet thickness,
the wavelength of the instability are all comparable,
the Weibel region moves to the $x$-direction
slightly slower than the TD ($V_{TD} \sim 0.65c$),
the local average velocity and the local fluid velocity differs,
and local plasma velocities depend on the distance from the TD.
However, the goal of our simple theory is to roughly understand the physics. 

\begin{figure}[thp]
\begin{center}
\ifjournal
\includegraphics[width={0.75\columnwidth},clip]{f4.eps}
\else
\includegraphics[width={0.75\columnwidth},clip]{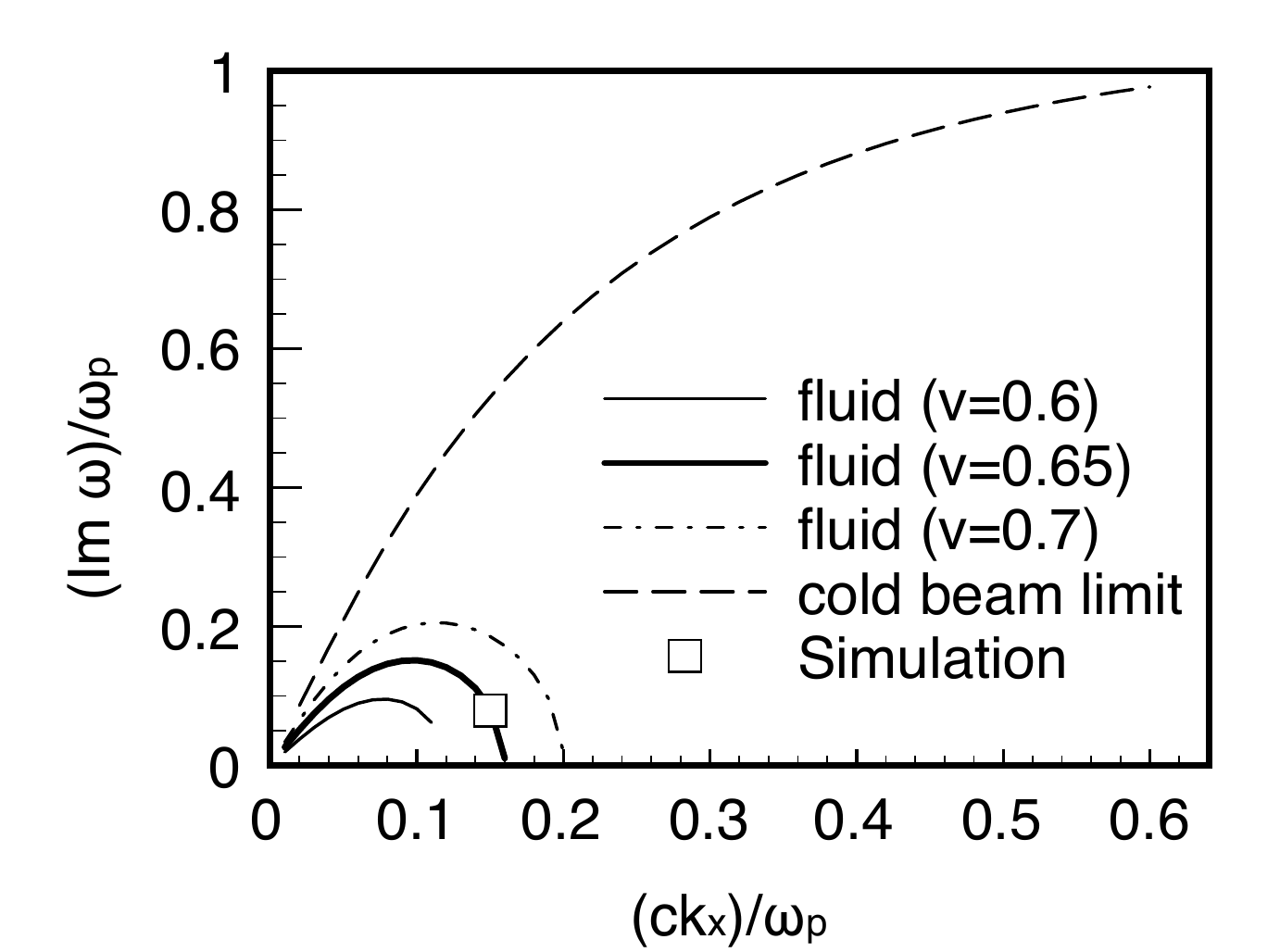}
\fi
\caption{
\label{fig:weibel}
Dispersion relation of
the two-dimensional purely-growing mode for $k_z=5k_x$.
The growth rate for three counter-streaming velocities ($v=0.6$,$0.65$,$0.7c$),
the cold-beam limit counterpart for $v=0.65c$ (\textit{dashed line}),
and observed rate (\textit{white square}) are presented.
}
\ifjournal
\includegraphics[width={\columnwidth},clip]{f5.eps}
\else
\includegraphics[width={\columnwidth},clip]{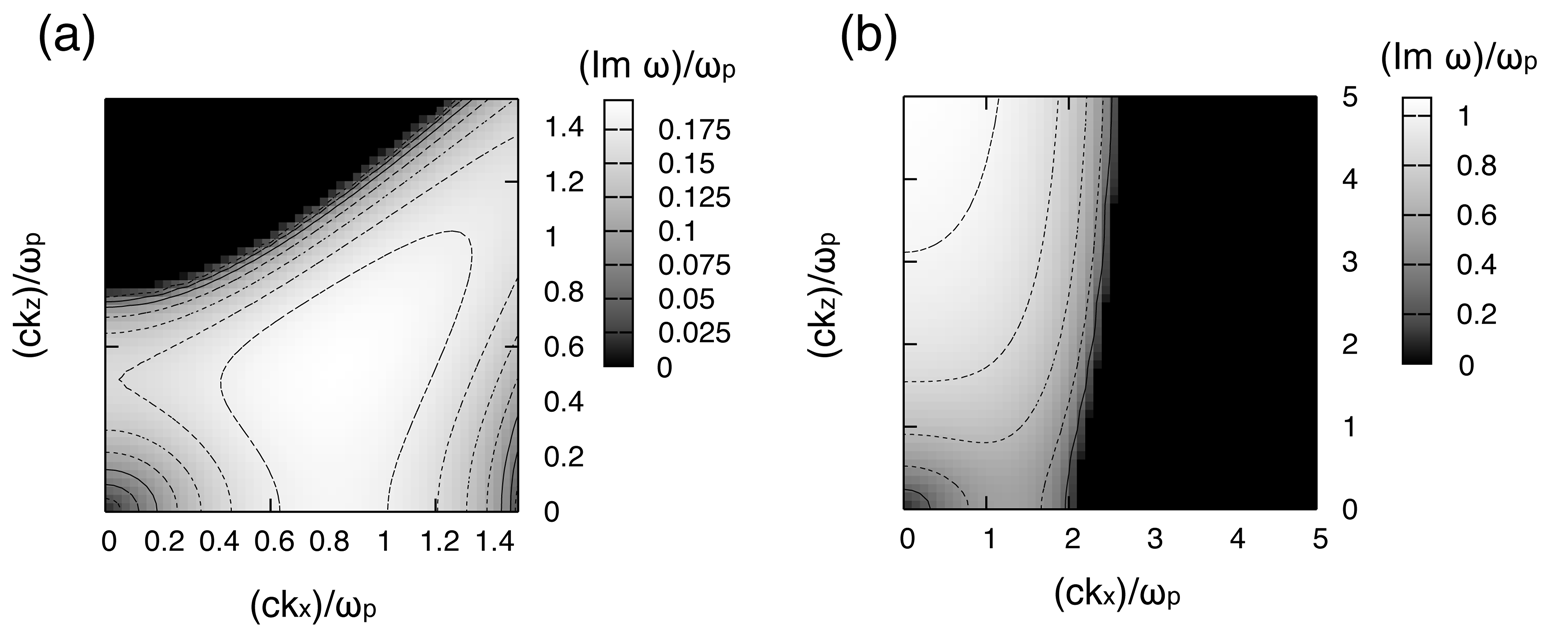}
\fi
\caption{
\label{fig:weibel2}
\textit{(a)}
Growth rate of the two-dimensional purely-growing mode
as a function of $\vec{k}=(k_x,k_z)$.
\textit{(b)} The same, but for the cold-beam limit case of $p=0$.
}
\end{center}
\end{figure}

Figure \ref{fig:weibel} shows
the dispersion relation of the two-dimensional mode for $k_z=5k_x$.
The linear analysis (\textit{bold line}) and the simulation data
are in good agreement. 
Further investigation shows that
the maximum growth rate is
on an order of $0.01$-$0.02\omega_p$
with the relativistic temperature of $T=mc^2$,
and that the cut-off (decline of the growth rate) is
rather sensitive to the counter-streaming velocity. 
The obtained mode is purely growing, and
it has an electromagnetic feature.
Because of the mathematical symmetry,
we obtain the other oblique modes for $(\pm k_x,\pm k_z)$
with the same growth rates.
Therefore, the two-dimensional rectangular structure is
obtained by superimposing these oblique modes.
The change separation structure (Figure \ref{fig:snap}\textit{c})
in the simulation frame can be explained by
the $z$-displacement by the instability.
It reflects both the density gradient inside the current sheet
and the Lorentz boost of the fast outflow streams.
The electrostatic component of the instability is relatively small.

We can also obtain
\newrev{the growth rate of the instability}
in counter-streaming cold beams \newrev{\citep{kazi98}}
by dropping the plasma pressure effect.
(One can remove pressure-related terms from eq. \ref{eq:fluid}
and employ $e_{sa} = n_{sa} m c^2$ instead of eqs. \ref{eq:p}.)
For comparison, the growth rate of the cold-beam limit is also presented
in Figure \ref{fig:weibel} (\textit{dashed line}). 
Obviously, the instability \newrev{grows substantially slower
than the cold-beam limit.}
One interpretation is that
imposing plasma pressure means the reduction of the anisotropy.
In a high temperature limit
where the four velocity of the counter-streams is relatively negligible,
the distribution becomes close to a single isotropic distribution.
Another interpretation is that
the relativistic pressure effect slows down \newrev{the growth rate},
as discussed in the relativistic studies
on the one-dimensional Weibel instability \citep{yoon87,yang93,yoon07}.
In relativistic temperature regime,
it is known that the Weibel instability is re-scaled by
Im $\omega \lesssim \omega_p/\hat{\gamma}^{1/2}$ in time
and $(\hat{\gamma}^{1/2} ck)/\omega_p$ in space,
where $\hat{\gamma}$ is the typical Lorentz factor of plasma maximum energy
\citep{yoon87}.
From the viewpoint of relativistic fluids,
the enthalpy term in equation \ref{eq:fluid} increases an effective inertia,
and then it slows down \newrev{the growth rate of the instability.}
In the present case, the term yields
$(\gamma_{sa}/c)^2(p_{sa} + e_{sa}) \sim (nm) \gamma_{sa}^2 ( 1 + [\Gamma/(\Gamma-1)] [T_{sa}/({mc^2})] )\sim 9n_{sa}m$.
Since it replaces the mass term inside the plasma frequency,
the instability in a relativistic hot plasma
grows slower than
the instability in the cold beam limit by a factor of $\sqrt{9}\sim 3$.
By comparing the enthalpy term
in relativistically hot limit ($\sim 4p/c^2$)
and 
in cold-beam limit ($\sim nm$),
we obtain the slow-down factor of ${p}^{1/2}$.
This is consistent with the scaling of the one-dimensional Weibel instability,
by a factor of $\hat{\gamma}^{1/2}$. 
In summary, the counter-streaming Weibel-type instability slows down
by the inertia effect of relativistic pressure.
Roughly speaking, the instability
is similarly re-scaled by a factor of $\hat{\gamma}^{1/2}$,
as the one-dimensional Weibel instability.

The panels in Figure \ref{fig:weibel2} present
growth rates of the obtained unstable modes
as a function of $\vec{k}=(k_x,k_z)$.
Both relativistic pressure case (Fig. \ref{fig:weibel2}\textit{a}) and
the cold-beam limit (Fig. \ref{fig:weibel2}\textit{b}) are shown.
The one-dimensional mode along $k_x=0$ is
the conventional Weibel instability,
which has electromagnetic features.
On the other hand, the mode along $k_z=0$ is
the electrostatic counter-streaming instability.
The typical mode in our simulation is $(k_x,k_z)=(0.15,0.75)$
with some amount of ambiguity.
As seen in Figure \ref{fig:weibel2}\textit{a}),
the obtained mode is rather close to
the one-dimensional Weibel instability. 
\newrev{It is important to note that the oblique mode grows
slightly faster than the one-dimensional Weibel instability,
and this is a signature of the counter-streaming Weibel-type instability.}
The central region and the right half
of Figure \ref{fig:weibel2}\textit{a} are
mainly occupied by the electrostatic-like mode.
Their growth rate is even faster, however,
since our theory depends on
the isotropic fluid pressure and the adiabatic condition (eq. \ref{eq:p}),
we think that our theory may be invalid,
especially in the short wavelength range of $(c/\omega_p) |\vec{k}| \gtrsim 1$.
In addition, in the high-pressure regime,
the two counter-streaming distributions overlap each other.
All these conditions are unfavorable to
describe the electrostatic modes in the parallel direction.
Meanwhile, the cold beam limit (Fig. \ref{fig:weibel2}\textit{b})
seems to be in good agreement with the \citet{saito}'s work,
based on \citet{kazi98}'s theory.
Their counter-streaming velocity $0.5c$ is comparable to ours of $0.65c$.



\section{Effect of the Weibel instability}

In this section, we investigate how the Weibel instability affects
the micro dynamics of plasmas motion and global dynamics of reconnection.
First we focus on the plasma motion near/in the downstream region,
because the Weibel instability occurs only in the downstream side of the TD.
Two characteristic regions will affect plasma motion ---
the TD and the Weibel region.
Before the Weibel instability appears,
particles in the downstream region are meandering in the current sheet.
Once they are hit or reflected by the TD,
they constantly travel into the $+x$-direction
because $x$-momentum is conserved
in the Harris current sheet configuration without $B_y$. 

\begin{figure}[thp]
\begin{center}
\ifjournal
\includegraphics[width={\columnwidth},clip]{f6.eps}
\else
\includegraphics[width={\columnwidth},clip]{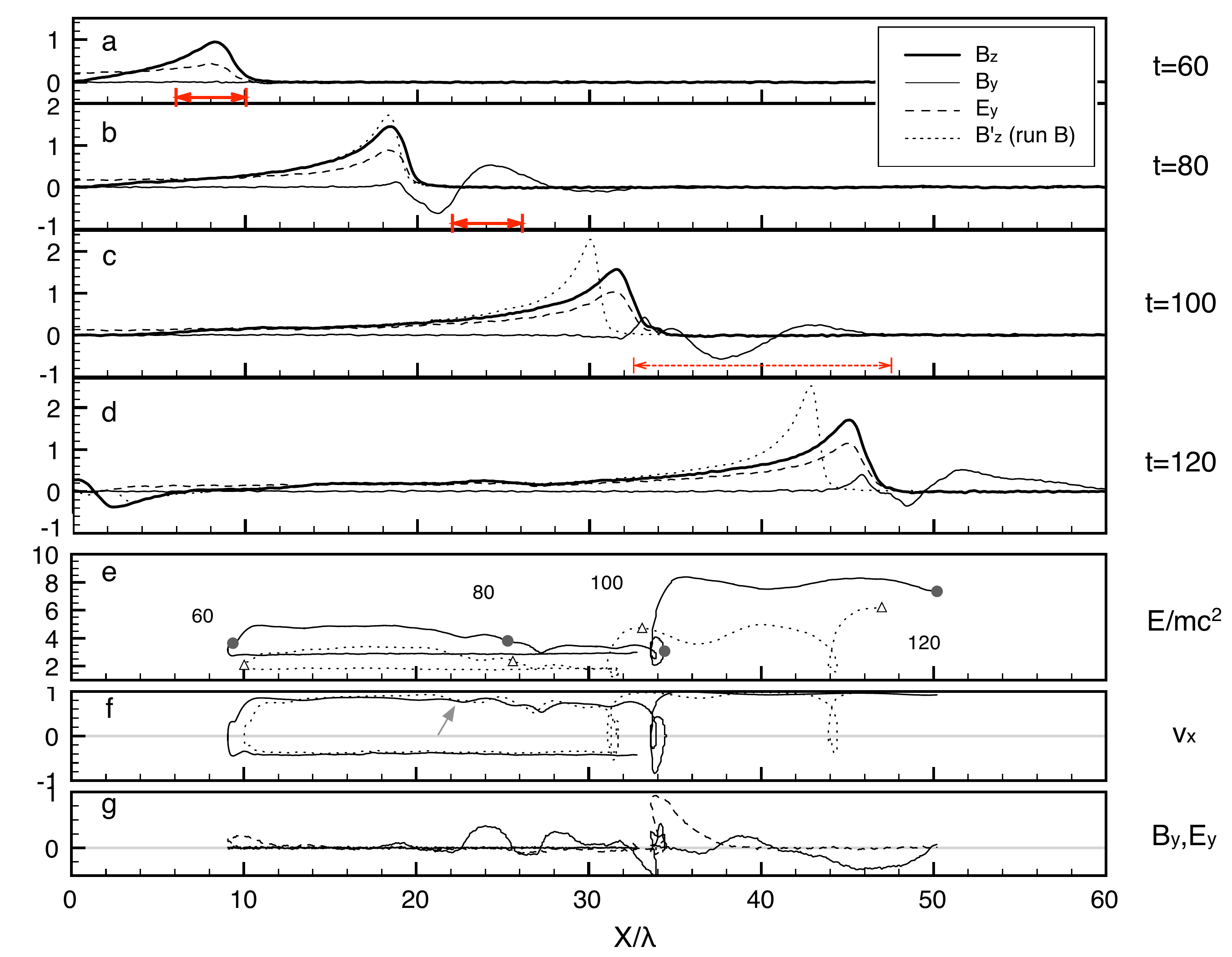}
\fi
\caption{
\label{fig:stack}
\textit{(a-d)} Field properties along the neutral plane ($z=0$);
$B_z/B_0$ (\textit{thick line}), $B_y/B_0$ (\textit{thin line}),
and $E_y/B_0$ (\textit{dashed line}) are presented.
The dotted line shows $B'_z/B_0$, obtained from a simulation without $B_y$.
\textit{(e-g)} Properties of selected particles;
\textit{(e)} energy,
\textit{(f)} velocity, and 
\textit{(g)} fields at its position are shown.
}
\end{center}
\end{figure}

\begin{figure}[thp]
\begin{center}
\ifjournal
\includegraphics[width={\columnwidth},clip]{f7.eps}
\else
\includegraphics[width={\columnwidth},clip]{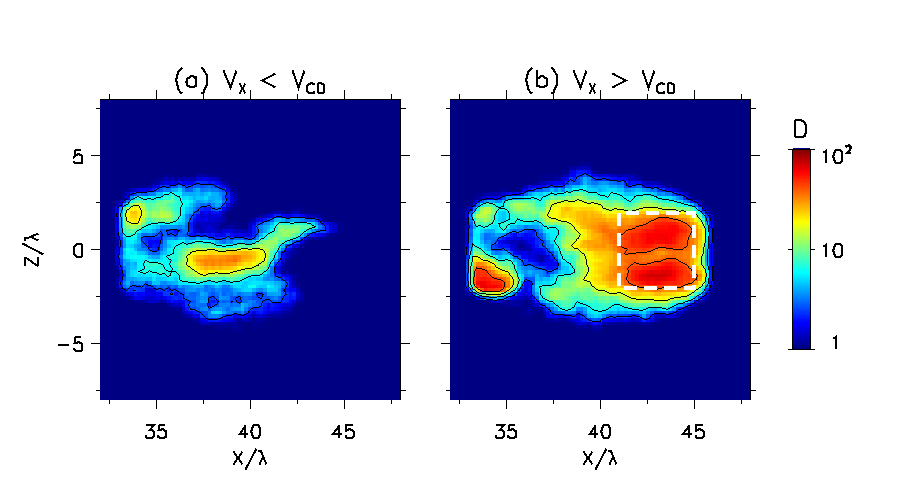}
\fi
\caption{(Color online)
\label{fig:dist}
\textit{(a)} Spatial distribution of selected positrons, whose $x$-velocity is slower than TD.
\textit{(b)} Spatial distribution of positrons, faster than TD.
}
\ifjournal
\includegraphics[width={\columnwidth},clip]{f8.eps}
\else
\includegraphics[width={\columnwidth},clip]{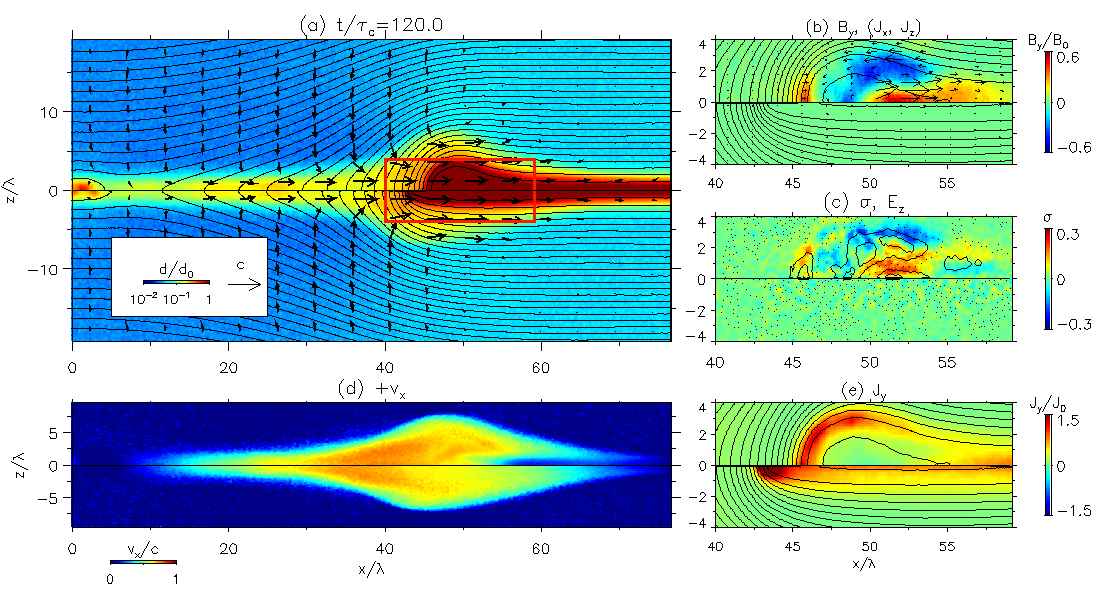}
\fi
\caption{(Color online)
\label{fig:t120}
\textit{(a-c)} Same as Fig. 1, but for $t/\tau_c=120$ in run A (\textit{upper half}) and in run B (\textit{buttom half}).
\textit{(d)} $x$-velocity and \textit{(e)} $y$-current at $t/\tau_c=120$ in run A (\textit{upper half}) and in run B (\textit{buttom half}).
}
\end{center}
\end{figure}

So, what happens after the Weibel magnetic fields $B_y$ appears?
In order to study plasma motion around the two characteristic regions,
we select $10^5$ super particles
($5 \times 10^4$ pairs)
that satisfies the following conditions;
they are 
(i) found in the piled-up region
($6\le x/\lambda \le 10$,$-2\le z/\lambda \le 2$)
at $t/\tau_c=60$
and (ii) found in the Weibel region
($22\le x/\lambda \le 26 $,$-2 \le z/\lambda \le 2$)
at $t/\tau_c=80$.
The $x$-ranges of these regions are
indicated by arrows in Figures \ref{fig:stack}\textit{a,b}.
Then, the spatial distribution of the selected particles are investigated.
The panels in Figure \ref{fig:dist} show
the distribution of the selected positrons at $t/\tau_c=100$.
The right panel shows the distribution of fast positrons.
The relevant $x$-range is indicated by the dashed arrow
in Figure \ref{fig:stack}\textit{c}.
The left panel presents slow positrons,
whose $x$-velocity $v_x$ is slower than that of the TD; $v_{TD}=0.65 c$.
The panel contains $10^4$ slow positrons (20\% of positrons). 
%

Roughly speaking, these positrons can be classified
into the following three groups. 
The first group is moving-away positrons,
who travels faster than the TD to the $+x$ direction.
The rectangle
(indicated by the white dashed line in Figure \ref{fig:dist}\textit{b})
is an approximate location of
the selected particles at $t/\tau_c=100$. 
After they are hit by the TD, they continue to escape into the $+x$ region,
faster than $V_{TD}$. 
The second group is found around $33<x/\lambda<40, z/\lambda \sim \pm 2$
in both two panels in Figure \ref{fig:dist}.
They are located along the magnetic field line,
which are connected to the TD. 
They have relatively small population, and
they do not always escape into the $+x$ direction.
We discuss the field-line modulation and
the current sheet expansion
later in this section. 
The last group is found along the neutral line ($z \sim 0$)
in Figure \ref{fig:dist}\textit{a}.
These particles are affected by the Weibel instability.
Their $z$-locations are positive around $40<x/\lambda<44$,
and negative around $36<x/\lambda<40$.
These $z$-displacements are due to the Weibel instability;
the effect of the out-of-plane field $B_y$
(Refer to Fig. \ref{fig:stack}\textit{c} for the polarity of $B_y$). 
The high density region near the TD ($x/\lambda <36, z/\lambda \sim 2$ in Figure \ref{fig:dist}\textit{a}) contains
both the second class of positrons along the field lines and
the third class of Weibel-affected positrons.
They are soon reflected by the TD, and then
we see their reflection in the other high density region near the TD
($x/\lambda <36, z/\lambda \sim -2$ in Figure \ref{fig:dist}\textit{b}).
We note that the magnetic field near the TD is not vertical,
but rather tilted into $+y$ direction,
because the TD hit the positive $B_y$ region at this time. 
Since 20\% of selected positrons are in Figure \ref{fig:dist}\textit{b},
the third group has relatively large population. 

We pick up $1.7 \times 10^3$ positrons from them,
which are found in the vicinity of the TD
($x/\lambda \le 33$, $-2 \le z/\lambda \le 2$) at $t/\tau_c=100$,
and then we examine their trajectories. 
The selection mainly consists of
the third class of Weibel-modulated positrons,
because the second class has fewer population.
Properties of two typical trajectories are shown
in Figure \ref{fig:stack}\textit{e,f} as a function of $x$.
We call them positron $A$ (\textit{solid line}) and
positron $B$ (\textit{dotted line}).
Throughout the simulation period ($0 \le t/\tau_c\le 120$),
they stay in the narrow region of $-2<z/\lambda<2$.
In Figure \ref{fig:stack}\textit{e},
marks show the particle position at the selected time stages
for comparison with Figures \ref{fig:stack}\textit{a-d}. 
Particle $A$ starts from $x/\lambda \sim 33$ in the $-x$ direction.
Its energy is originally $\varepsilon \sim 3 mc^2$.
Around $t/\tau_c \sim 60$, it collides with the TD and then
it turns its way to the $+x$ direction.
Its energy increases to $\varepsilon \sim 5 mc^2$
via the interaction with the TD.
Figure \ref{fig:stack}\textit{g} show the field properties
$B_y$ (\textit{solid line}) and $E_y$ (\textit{dashed line})
at positron $A$'s position. 
Near $x/\lambda \sim 22$,
it feels relatively strong $B_y$ in the Weibel-active region,
and then its $x$-momentum is transported to $z$ momentum
(also, to $y-$momentum through the meandering motion).
Consequently, its $x$-velocity starts to slow down,
as indicated by the arrow in Figure \ref{fig:stack}\textit{f}.
Since the TD travels relatively fast ($v_{TD} \sim 0.65c$),
the TD eventually catches up positron $A$ and hits it again.
This time, the positron $A$ gains more energy by the motional electric field $E_y$,
because more magnetic fields are piled up than during the first impact.
Now its energy goes up to $\varepsilon/mc^2 \sim 8$. 
So, the Weibel field slow down the escaping particles, and then
it enables multiple interactions with the TD.
We even find three-time or four-time interactions with the TD
in selected positron trajectories. 
For example, positron $B$ (\textit{dotted line}) are hit
by the TD twice ($x/\lambda \sim 31, 44$)
after the first interactions with the TD at $x/\lambda \sim 9$.
If we study further long-time evolution,
these positrons will be hit by the TD multiple times.

Interestingly, it seems that
only low-energy particles are reflected by the Weibel region;
high-energy particles are insensitive to the Weibel fields and then
they easily escape to the $+x$ direction,
when their energy $\varepsilon/mc^2$ exceeds $7$-$8$. 
The typical kinetic energy gain by the TD reflection
ranges up to $\varepsilon/mc^2 \sim 10$-$20$,
due to enhanced pileup electric field.
Therefore, among the reflected particles,
low-energy particles are trapped between the TD and the Weibel region,
and then they are heated by the multiple interactions with the TD.
They can not escape into the outflow region,
until they become energetic enough --- 
their gyro radii exceeds the scale of the Weibel structure
($\gamma c/\omega_c \gg \gamma^{1/2} c/\omega_p$).
So, this result indicates that
the Weibel instability enhances plasma heating in the downstream side of the TD,
while it is not likely to enhance high-energy particle acceleration.
The threshold energy will increase
as the system condition becomes more relativistic,
because the electron skin depth is relatively larger by default,
and because the typical scale of the Weibel instability
becomes even larger by a factor of $\hat{\gamma}^{1/2}$ or $p^{1/2}$.

Next, in order to study the Weibel mode effect
to global reconnection structure,
we carried out another simulation.
The new run (run B) starts from
intermediate data of run A at $t/\tau_c=60$,
and then we artificially reduce $B_y$ to $0$ in run B.
The quadrupolar magnetic fields near the reconnecting region
\citep{hesse94,oe01,nagai01} do not appear
in pair plasma reconnection,
because Hall physics depends on different ion and electron masses.
In antiparallel configurations,
magnetic reconnection involves $B_x$ and $B_z$, and
only the Weibel instability or anisotropy-driven instabilities 
generates $B_y$.
Because of the artificial reduction of $B_y$,
the system slightly loses energy.
\newrev{The total energy in run B is smaller than the total energy in run A
by $0.15\%$ at the end of the simulation ($t/\tau_c=120$).
This energy loss is much smaller than typical kinetic, magnetic,
and internal energies.}
The panels in Figure \ref{fig:t120} show
the late-time evolution of the two runs at $t/\tau_c=120$.
For comparison, the upper halves present the results of run A,
while the lower halves present those of run B.
\newrev{At this time, there should be no substantial boundary effects,
since the evolution time of the reconnection region is smaller than 
a wave propagation time to the $z$-periodic boundaries.
In addition, the magnetic field lines at the $X$-type region
come from $z/\lambda\sim \pm 15$ around $x/\lambda = \pm 76.8$
as seen in Figure \ref{fig:t120}\textit{a}.
The reconnection still involves magnetic flux
inside the main simulation domain.}

One can see the difference
in global structure in Figure \ref{fig:t120}\textit{a}.
The current sheet seems to be broadened in run A,
while it remains thin in run B.
The position of the TD front differs, too.
The TD is located at $x/\lambda \sim 46$ in run A.
On the contrary, the TD is located at $x/\lambda \sim 43$ in run B
--- the TD front can not penetrate into the $x$-direction as run A.
In Figures \ref{fig:stack}\textit{a-d},
the $B_z$ profile along the neutral line in run B
is also presented by a dotted line.
The peak plasma density at the downstream of the TD is
$d/d_0 \sim 2.4$ in run A,
while plasmas are much compressed near the TD; $d/d_0 \sim 5$ in run B.
These differences can be explained by
the current sheet expansion by the Weibel instability.
In run A, the Weibel instability transfers some of plasma $x$-momentum
into $z$-momentum (and also $y$-momentum via meandering motion).
Therefore, it reduces plasma $x$-pressure in the downstream region,
and the TD front can move further distance into $x$-direction.
The increased $z$-momentum leads to the current sheet expansion,
and then magnetic field lines become more round.
The current sheet continues to expand
as long as plasma $z$-momentum is continuously supplied
through the Weibel activity,
from reconnection outflow in the upstream region.
On the contrary, plasmas and anti-parallel field lines are
rather confined near the current sheet in run B.
The TD can not penetrate into the outflow direction as run A,
therefore,
the maximum amplitude of the pileup field is stronger
(See Figs. \ref{fig:stack}\textit{a-d} for the $B_z$ profiles in run B)
and 
the inflow speed near the TD is slightly slower than in run A
(Fig. \ref{fig:t120}\textit{d}). 

The total reconnected flux $\Sigma|B_z|$
along the neutral plane is the same in runs A and B,
because it is controlled by the physics of the upstream region;
magnetic reconnection near the $X$ points. 
Around $(x,z)\sim(0,0)$, we observe
a small magnetic loop in the current sheet.
\newrev{This is not a projection of the $X$-point,
but a secondary magnetic island},
which appears after $t/\tau_c \sim 105$.
Its formation mechanism is unclear.
Since these secondary islands are also found in
non-relativistic reconnection in pair plasmas \citep{dau07},
the island formation will be common feature
in a low-density current sheet in large scale simulations.
Anyway, 
we do not see noticeable difference ---
both run A and run B are almost same in the upstream region.
\newrev{At this stage the Weibel activity is not likely affecting
the central reconnection region,
because information can not have widely spread over
the central reconnection region.
For example, an information of large $B_y$-perturbation at $x/\lambda \sim 20$
at $t/\tau_c=80$ (Fig. \ref{fig:linear}) can not 
arrive at the $X$-type region before $t/\tau_c = 100$. 
On a longer time scale, 
the Weibel instability may have an impact on the reconnection rate
since it changes the magnetic field line topology
by expanding the current sheet \cite{swisdak}.}

In the case of run A,
Figure \ref{fig:t120}\textit{b} and 
Figure \ref{fig:t120}\textit{c} show
the downstream field structure in more detail.
In Figure \ref{fig:t120}\textit{b},
the magnetic field lines near $x/\lambda \sim \pm 1$
at the right boundaries
are set to connect to $x/\lambda = \pm 1$ at periodic boundaries.
The field line shifts to $z/\lambda \sim 3$ at the thickest point
due to the current sheet expansion in run A,
while the field line stays around $z/\lambda \sim -1$ in run B.
In this stage, the Weibel instability is also active
outside the neutral line around $z/\lambda \sim 2$,
as well as along the neutral line.
The $xz$ current system is well developed around the $B_y$ regions.
Charge distribution (Fig. \ref{fig:t120}\textit{c}) is
correlated to the $x$-current system (Fig. \ref{fig:t120}\textit{b});
we see positron-rich $J_x>0$ region and electron-rich $J_x<0$ region.
Compared with the early stage in Figure \ref{fig:snap},
these structures are rather elongated into $x$-direction.
This is consistent with many studies on Weibel instability;
elongated ``filament'' structure or current channels are
commonly observed in well-developed stage of the Weibel instability. 
In run B, there is no current system in the $xz$ plane.
Regarding the outflow structure in Figure \ref{fig:t120}\textit{d}, 
plasma flow is rather bifurcated in run A,
due to the $z$-displacement of plasmas.
We can see a significant difference in the $J_y$ current structure
in Figure \ref{fig:t120}\textit{e}.
In run A, the current region is located in front
of the broadened plasma region.
The energy conversion $\vec{J}\cdot\vec{E}$ mainly takes place 
in the vertical current front there.
In run B, the $y$-current structure is enhanced 
around the small spot near the TD,
and then energy conversion takes place there.

Regarding the composition of the energy in the system of interest,
two runs slightly differ in accordance with the field line topology;
the summary of the upstream reconnection field energy
$\Sigma{(B_x^2+B_z^2+E_y^2)/8\pi}$ is almost same in both two runs,
but run A has more field energy (120\%) in $\Sigma{(B_z^2+E_y^2)/8\pi}$
compared with run B. 
The total plasma kinetic energy
$\Sigma{(\gamma-1)mc^2}$ is almost same, however,
in run A, a slight percentage of them ($0.5\%$) are converted again
into the Weibel-related field energy; $\Sigma{(B_y^2+E_x^2+E_z^2)/8\pi}$.
The total amount of Weibel-related field energy is equivalent to
$\sim 5.3 \lambda^2 (B_0^2/8\pi) $.
This is substantially smaller
($10^{-1}$-$10^{-2}$) than that of the upstream-related field energies,
because the Weibel active region is relatively small.

\begin{figure}
\begin{center}
\ifjournal
\includegraphics[width={0.9\columnwidth},clip]{f9.eps}
\else
\includegraphics[width={0.9\columnwidth},clip]{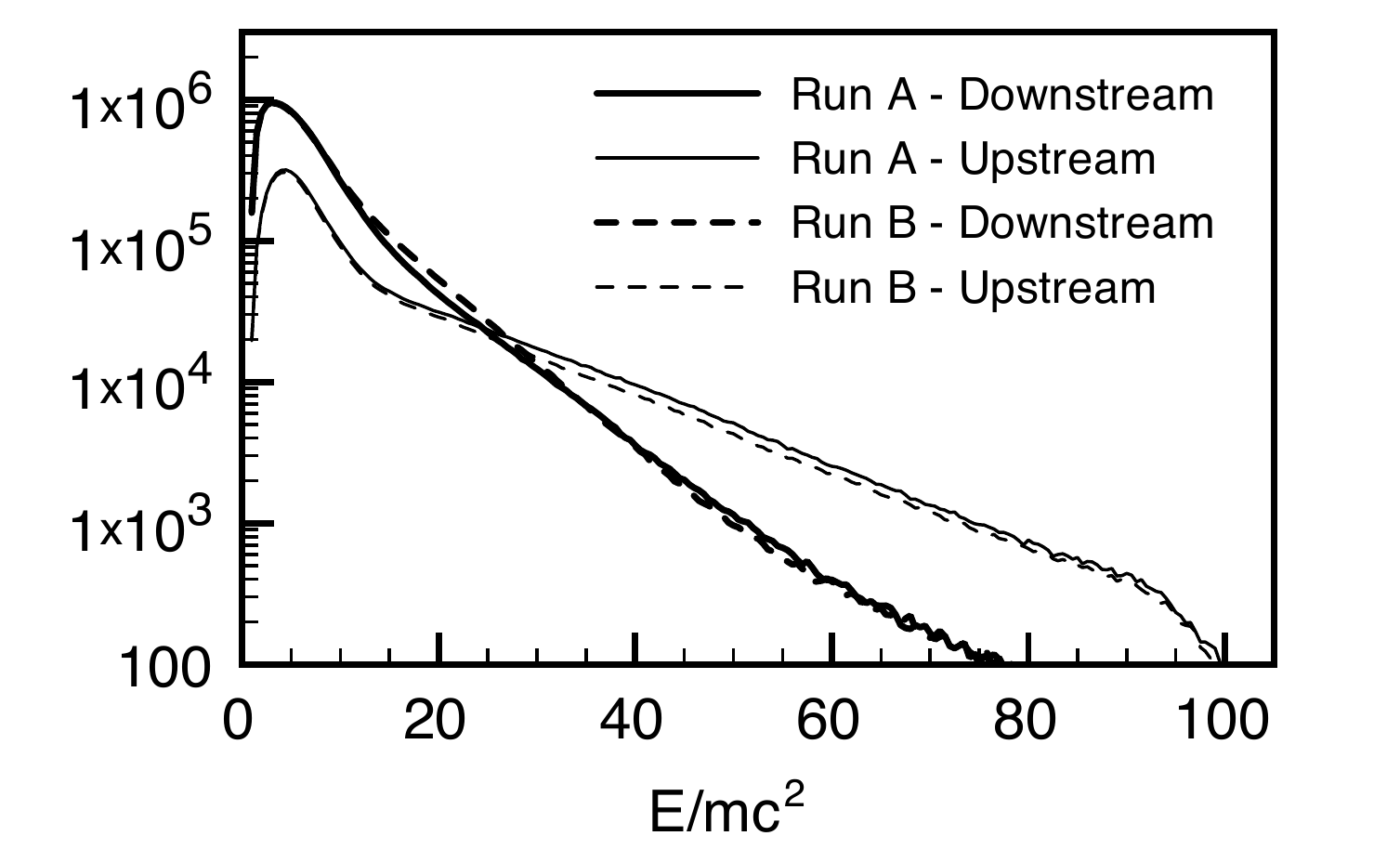}
\fi
\caption{
\label{fig:espec}
Energy spectra in the right half main simulation domain.
}
\end{center}
\end{figure}

Figure \ref{fig:espec} presents
energy spectra in the regions of interest.
The spectra of two runs look similar, too.
However, in order to distinguish the difference clearer,
these spectra are divided into two parts by the TD;
in the downstream region of the TD,
and in the upstream region of the TD.
Note that the TD is located in the further downstream in run A.
We observe a high-energy nonthermal tails in their spectra
in the upstream side.
This is due to $dc$ particle acceleration or piled-up acceleration
in the upstream side \citep{zeni01,zeni07a}. 
In the mid-energy range ($20\lesssim \varepsilon/mc^2 \lesssim 60$),
run A has slightly more high-energy population.
We think this is due to the larger volume of the upstream region.
Since particles can stay longer
inside the larger upstream region,
or the main site of particle acceleration,
more particles are accelerated into high energy range in run A.
In the low-energy range around $\varepsilon/mc^2 \sim 10$,
We expected that plasma heating is enhanced through
multiple interaction by the TD in run A,
but the enhancement is too small (even in linear scaling).
One reason is that the Weibel region is too small.
Furthermore, the Weibel region not only reflects the escaping particles,
but also it hits the pre-existing particles.
So, as a result, the net effect will be small.
Meanwhile, in run B, plasma population is slightly enhanced
around $\varepsilon/mc^2 \sim 20$.
It is difficult to discuss this energy range,
because too many effects are relevant.

\section{Discussions and Summary}

In ion-electron reconnection,
it is well known that quadropolar out-of-plane fields $B_y$
appears in the vicinity of the $X$-type region
\citep{hesse94,oe01,nagai01}. 
However, quadropolar structure disappears
in pair plasmas with an equal temperature \citep{bessho07}. 
In the present case, the out-of-plane fields are found
in the downstream of the reconnection outflow region.
\citet{dau07} reported similar structure
in non-relativistic pair plasma reconnection, and
they argued that it comes from some type of firehose instability.
Since both the firehose instability and the Weibel instability
belong to the anisotropy-driven instabilities,
the generation of the out-of-plane fields in the downstream
would be common feature in pair plasma reconnection. 
We identified that
the out-of-plane field $B_y$ is generated
by the Weibel instability downstream of the TD. 
A comparison with another run (with $B_y$ artificially suppressed)
demonstrates that 
the Weibel instability leads to a significant modulation
of the downstream structure; further penetration of outflows,
the current sheet expansion and the bifurcated downstream jets.
We expect that
the formation of ``T-shaped current sheet'' \citep{claus04}
can be explained by the current sheet expansion
by the Weibel instability.

In three dimensions,
the Weibel instability also generates
the vertical magnetic field $\pm \delta B_z$ and
the out-of-plane electric field $\pm \delta E_y$.
\newrev{The Weibel instability will lead to
a filament-like development of the $x$-currents,
involving small-scale reconnection of
perturbed magnetic field lines \citep{cal01}.
The plasma's $x$-momentum will be transferred
to $y$-momentum as well as to $z$-momentum.
We expect that
the TD penetrates further downstream into the outflow region,
because plasma $x$-pressure will be more efficiently scattered.
Meanwhile, the current sheet expansion in the downstream region
may be less apparent,
because all of the $x$-momentum is not transformed into $z$-momentum.
Regarding the particle acceleration,
we will observe more high-energy particles,
because the TD will further penetrate into downstream and then
the upstream acceleration site expands.
In addition, high-energy particles from the upstream region
may also be affected by the Weibel fields.
In two dimensional case, such high energy particles are
rather insensitive to the Weibel region, and 
low-energy reflected particles bounce
between the TD and the Weibel-active region.
However, in three dimensions,
the Weibel magnetic field can affect high energy particles,
especially when it is antiparallel ($B_z<0$) to the pile-up field.
Thus, some higher-energy particles may bounce
between the TD and the Weibel-active region.
Along with the expansion of the upstream acceleration site,
particle acceleration is likely to be enhanced. }

Furthermore, we should consider
all other instabilities in three dimensions.
It is known that the relativistic drift kink instability (RDKI)
quickly modulates the current sheet in a relativistic pair plasmas
\citep{zeni05a,zeni07a}.
Its typical growth rate in this configuration is
$\tau_c\omega_i \sim 0.1$,
while $\tau_c\omega_i \sim 0.03$ for the tearing instability.
Although the RDKI grows slower than the Weibel instability,
the RDKI is a macro instability,
and it may inhibit the reconnection process
by modulating the current sheet \citep{zeni05b},
while the Weibel instability is
the sub-product of the reconnection outflow. 
Since the RDKI slowly widens the current sheet,
unmagnetized or weakly magnetized region becomes wider.
Therefore, we expect that the Weibel instability is active
in a wider region inside the modulated current sheet.
\newrev{In addition, since the Weibel instability involves $y$-structure,
repeated collision between the TD and the Weibel fields may lead to
the instability of the TD front in the $xy$ plane
(e.g. the interchange instability of the reconnection jet front \citep{naka02}).
The Weibel instability in three dimensions
will be an interesting problem to challenge. }

The Weibel instability will also occur
under non-relativistic temperature condition of $T \ll mc^2$.
In this regime, usually the electron skin depth becomes
smaller than the other scales like electron gyro radius, and so
the Weibel instability occurs
in a shorter time/spatial scale in reconnection. 
As long as it occurs inside reconnection outflow structure,
the physics will be the same. 

In ion-electron plasmas,
the Weibel instability will work for electrons,
and then it may contribute to quick electron heating. 
Although it is not clear whether
the sharp TD is formed in the outflow region in ion-electron plasmas,
the multiple interaction with the TD will also be possible. 
Similarly, enhanced heating may also occur
near the fast shock or the other discontinuities.
\newrev{
In solar cases, it is reported that
hard X-ray emission comes from the small spot
near the loop top of magnetic field lines \citep{masuda94}.
Shock-related electron heating may take place
in the downstream of reconnection outflow, and
the Weibel instability and the relevant bounce effect
may contribute to the enhanced electron heating there.
In addition, the Weibel instability may play a role in quick electron heating
inside the nanoflare jets, in the context of coronal heating problem.
}

In more generalized configuration of magnetic reconnection,
for example, in magnetic reconnection with uniform guide field $B_y$,
the situation will differ substantially.
Since the guide field $B_y$ scatters $x$-momentum into $z$-momentum,
the wavevector of the Weibel instability is likely to be in the $y$ direction. 
However, since the outflow is slower than the antiparallel case,
and since the ambient magnetic field $B_y$ stabilizes the instability,
the Weibel instability will be less active. 
The situation will be more complicated in relativistic pair plasmas, 
because charge neutrality often breaks down
in the outflow region \citep{zeni07b}. 
Therefore, how plasma anisotropy disappears in the guide field case
remains to be solved. 

On the viewpoint of energetics,
the ultimate energy source of the Weibel instability is
the plasma bulk energy of the reconnection jet,
which is expelled by the magnetic energy in the inflow region.
Initial magnetic energy is converted to
plasma energy of reconnection jet, and
partially to magnetic energy of the Weibel region.
Then, the Weibel activity modifies the downstream reconnection structure,
which potentially changes the downstream energy conversion process once again!

Finally, let us briefly summarize this paper.
We investigated the role of the Weibel instability in the reconnection context.
We demonstrated the following new results;
(1) the Weibel instability occurs in the downstream of the reconnection outflow,
(2) the counter-streaming Weibel instability is also affected by the relativistic pressure effect,
and (3) the Weibel instability significantly modifies the downstream reconnection structure.  Since the Weibel instability is a micro process, it may play a role in various macro instabilities, such as magnetic reconnection, the RDKI, the Kelvin-Helmholtz instability, as well as collisionless shocks.


\begin{acknowledgments}
The authors enjoyed fruitful discussions with K. Schindler, M. Kuznetsova,
S. Saito, M. Swisdak and Y. Liu. 
This research was supported by facilitates of JAXA
and the NASA Center for Computational Sciences.
\end{acknowledgments}

\clearpage

\clearpage

\clearpage

\clearpage

\clearpage

\end{document}